\documentclass[copyright,creativecommons]{eptcs}
\providecommand{\event}{Infinity 2013}
\usepackage{amssymb,amsmath,times,amsfonts,times,color,graphicx,latexsym,url}
\newtheorem{Algorithm}{Algorithm}

\def    \to     {\rightarrow}

\def    \l      {\lambda}

\def    \R      {{\cal R}}

\def    \ni     {\noindent}

\def    \se     {\subseteq}

\def    \lcc#1  {\langle\!\langle #1 \rangle\!\rangle}
\def    \rcc#1  {\langle #1 \rangle}
\def    \cc#1   {[#1]}

\def    \ft#1   {\footnote{#1} }

\def	\nodeq#1 {\stackrel{#1}{\sim}}
\def	\trans#1 {\stackrel{#1}{\longrightarrow}}

\def	\comment#1 {}

\def    \comm#1  {\begin{center} {\tt $<<<$ #1 $>>>$ } \end{center}}

\def\GF{\hbox{\raise 1.6pt \hbox{$\varphi$}}}

\newcommand \ov[1]{\overline{#1}}

\renewcommand{\R}{{\mathbb{R}}}

\newcommand{\tp}{\tilde{P}}
\newcommand{\tv}{\tilde{V}}
\newcommand{\sig}{\mathcal{T}}

\begin{document}

\title{Algorithmic Verification of Continuous\\ and Hybrid Systems}
\author{Oded Maler \\
CNRS-VERIMAG, \\
University of Grenoble \\
 France}

\def\titlerunning{Algorithmic verification of continuous and hybrid systems}
\def\authorrunning{O. Maler}

\maketitle

\begin{abstract}
We provide a tutorial introduction to \emph{reachability computation}, a  class of computational techniques that exports verification technology toward continuous and hybrid systems. For open under-determined systems, this technique can sometimes replace an infinite number of simulations.
\end{abstract}

\section{Introduction}

The goal of this article is to introduce the fundamentals of algorithmic verification of \emph{continuous dynamical systems} defined by \emph{differential equations} and of hybrid dynamical systems defined by \emph{hybrid automata} which are dynamical systems that can switch between several modes of continuous dynamics. There are two types of audience that I have in mind. The first is verification-aware computer scientists who know finite-state automata and their algorithmic analysis. For this audience, the conceptual scheme underlying the presented algorithms will be easy to grasp as it is mainly inspired by symbolic model-checking of non-deterministic automata. Putting aside dense time and differential equations, dynamical systems can be viewed by this audience as a kind of infinite-state reactive program defined over the so-called real numbers. For this reason, I will switch from continuous to discrete-time discourse quite early in the presentation.

 The other type of audience is people coming from control or other types of engineering and applied mathematics. These are relatively well versed in the concrete mathematics of continuous systems and should first be persuaded that the verification question is interesting, despite the fact that airplanes can fly (and theoretical papers can be written) without it. In my attempts to accommodate  these two types of audience, I have written some explanation that will look trivial to some but this cannot be avoided while trying to do genuine inter-disciplinary research\ft{The need to impress the members of one's own community is perhaps the main reason for the sterility of many attempts to do inter-disciplinary research.} (see also \cite{unified} for an attempt to unify discrete and continuous systems and \cite{amir-dawn} for some historical reflections). My intention  is to provide a synthetic introduction to the topic rather than an exhaustive survey, hence the paper is strongly biased toward techniques closer to my own research.  In particular, I will not deal with decidability results for hybrid systems with simplified dynamics and not with deductive verification methods that use invariants, barriers and Lyapunov functions. I sincerely apologize to those who will not find citations of their relevant work. 

The rest of the paper is organized as follows. Section~2 describes the problem and situates it in context. Section~3 gives the basic definitions of reachability notions used throughout the paper. Section~4 presents the principles of set-based computation as well as basic issues related to the computational treatment of sets in general and convex polytopes in particular.  Section~5 is devoted to reachability techniques for \emph{linear} and affine dynamical systems in both discrete and continuous time, the domain where a lot of progress has been made in recent years. The extension of these techniques to hybrid and non-linear systems, an active area of research, is discussed in Section~6.  I conclude with some discussion of related work and new research directions.

\section{The Problem}

This paper is concerned with the following problem that we define first in a quasi-formal manner:

 \emph{Consider a continuous dynamical system with input defined over some bounded state space $X$ and governed by a differential equation of the form\ft{We use the physicists' notation where $\dot{x}$ indicates $dx/dt$. It is amusing to note that the idea of having a dedicated notation for the special variable called Time, is not unique to Temporal Logic.}   $$\dot{x}=f(x,v)$$ where $v[t]$ ranges, for every $t$, over some pre-specified bounded set $V$ of admissible input values. Given a set $X_0\subset X$, compute all the states visited by trajectories of the system starting from any $x_0\in X_0$.}

The significance of this question to control is the following: consider a controller that has been designed and connected to its plant and which is subject to external disturbances modeled by $v$. Computing the reachable set allows one to verify that all the behaviors of the closed-loop system stay within a desired range of operation and do not reach a forbidden region of the state space. Proving such properties for systems subject to uncontrolled interaction with the external environment is the main issue in verification of programs and digital hardware from which this question originates. In verification you have a large automaton with inputs that represent non-deterministic (non-controllable) effects such as behaviors of users and interactions with other systems and you would like to know whether there is an input sequence that drives the automaton into a forbidden state.

Before going further, let me try to situate this problem in the larger control context. After all, control theory and practice have already existed for many years without asking this question nor trying to answer it. This question distinguishes itself from traditional control questions in the following respects:
\begin{enumerate}
\item It is essentially a {\em verification} rather than a {\em synthesis} question, that is, the controller is assumed to exist already. However, it has been demonstrated that variants of reachability computation can be used for synthesizing switching controllers for timed \cite{AMP,AMPS,tiga} and hybrid \cite{procieee} systems.

\item External disturbances are modeled \emph{explicitly} as a {\em set} of  admissible inputs, which is not the case for certain control
formulations.\footnote{See \cite{adversary} for a short discussion of this intriguing fact.}  These disturbances are modeled in a \emph{set-theoretic} rather than \emph{stochastic} manner, that is, only the set of \emph{possible} disturbances is specified without any probability induced over it. This makes the system in question look like a system defined by \emph{differential inclusions} \cite{aubin} which are the continuous analog of non-deterministic automata: if you project away the input you move from $\dot{x}=f(x,v)$ to $\dot{x}\in F(x)$. Adding probabilities the the inputs will yield a kind of a stochastic differential equation \cite{astrom1970}.

\comment{
I think there are indeed 5 areas to mention
briefly: 1) PID (99\% of real control; 2) Kalmanism (nice linear
theory); 3) Optimal control and differential games; 4) Robust control
and 5) Model-predictive - something relatively more computational. For
example, in the linear Kalmanistic theory disturbance is modeled
implicitly as something that takes the system away from the
equilibrium point, to which the control takes it back, without a
disturbance. In robust control disturbances for linear systems are
modeled in the frequency domain. What is assumed and modeled in PID
and in optimal control. All this can be updated due to new insights
gained.}

\item The information obtained from reachability computation covers also the {\em transient behavior} of the system in question, and not only its \emph{steady-state} behavior. This property makes the approach particularly attractive for the analysis of {\em hybrid} (discrete-continuous, numerical-logical) systems where the applicability of analytic methods is rather limited. Such hybrid models can express, for example, deviation from idealized linear models due to constraints and saturation as well as other switching phenomena such as thermostat-controlled heating or gear shifting, see \cite{ejc} for a lightweight introduction to hybrid systems and more elaborate accounts in books, surveys and lecture notes such as \cite{schaft,morari,lygeros2001art,johansson2002piecewise,liberzon2003switching,tripakis2009modeling,tabuada2009verification,dang2009tools,cassandras2010stochastic,rajeev-emsoft}.

\item The notion of {\em to compute} has a more effective flavor, that is, to develop algorithms that produce a representation of the set of reachable states (or an approximation of it) which is computationally usable, for example, it can be checked for intersection with a bad set of states.

\end{enumerate}

Perhaps the most intuitive explanation of what is going on in reachability computation (and verification in general) can be given in terms of {\em numerical simulation}, which is by far the most commonly-used approach for validating complex systems. Each individual simulation consists of picking {\em one} initial condition and {\em one} input stimulus (random, periodic, step, etc.), producing the corresponding trajectory using numerical integration and observing whether this trajectory behaves properly. Ideally, to be on the safe side, one would like to repeat this procedure with \emph{all} possible disturbances which are uncountably many. Reachability computation achieves the same effect as {\em exhaustive} simulation by exploring the state space in a ``breadth-first'' manner: rather than running each individual simulation to completion and then starting a new one, we compute at each time step all the states reachable by \emph{all} possible one-step inputs from states reachable in the previous step (see \cite{cfromcs} for a more elaborate development of this observation and \cite{under-det} for a more general discussion of \emph{under-determined} systems and their simulation). This set-based simulation is, of course, much more costly than the simulation of an individual trajectory but it provides more confidence in the correctness of the system than a small number of individual simulations would. The paper is focused on one popular approach to reachability computation based on discretizing time and performing a kind of set-based numerical integration. Alternative approaches are mentioned briefly at the end.

\comment{The set-based numerical integration approach on which we focus in this paper
is the most developed so
 is based on
}

\comment{general assessment of this approach and mention some related work and
alternative techniques.}

\section{Preliminaries}

\newcommand\inp{\zeta}
\renewcommand\trans[1]{\stackrel{#1}{\longrightarrow}}
\renewcommand\ov[1]{\overline{#1}}
\newcommand\ul[1]{\underline{#1}}
\newcommand\class{\mathcal{C}}
\newcommand\op{\circ}
\renewcommand\l{\lambda}

\begin{figure}
\begin{center}
\scalebox{0.45}{ 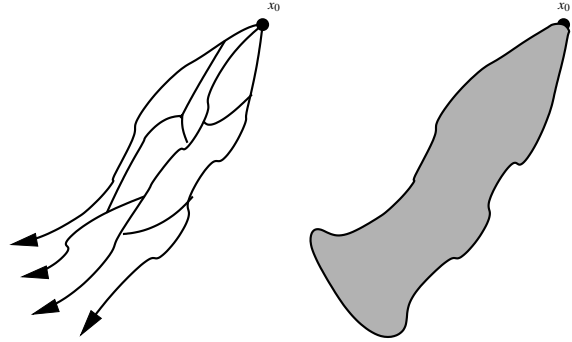 }
\caption{\label{fig:cont-tree} Trajectories induced by input signals from $x_0$ and the set
of reachable states.}
\end{center}
\end{figure}

We assume a time domain $T=\R_+$ and a state space $X\se \R^n$.
A trajectory is a measurable partial function $\xi:T \to X$ defined
over all $T$ (infinite trajectory) or over an interval $[0,t]\subset T$
(a finite trajectory).  We use the notation $\sig(X)$ for all such
trajectories and $|\xi|=t$ to denote the length (duration) of finite
signals. We consider an input space $V\se \R^m$ and likewise use $\sig(V)$
to denote input signals $\inp :T \to V$. A continuous dynamical system $S=(X,V,f)$ is a
system defined by the differential equation
\begin{equation}
\label{eq:dynamics}
\dot{x}=f(x,v).
\end{equation}
We say that $\xi$ is the {\em response} of $f$ to $\inp$ from $x$ if $\xi$
is the solution of (\ref{eq:dynamics}) for initial condition $x$ and
$v(.)=\inp$. We denote this fact by $\xi=f_x(\inp)$ and also as
$$x\trans{\inp/\xi}x'$$ when $|\inp|=t$ and $\xi[t]=x'$.  In this case
we say that $x'$ is reachable from $x$ by $\inp$ within $t$ time and
write this as
$$R(x,\inp,t)=\{x'\}.$$
This notion speaks of {\em one} initial state,
{\em one} input signal and {\em one} time instant and its
generalization for a {\em set} $X_0$ of initial states, for {\em all}
time instants in an interval $I=[0,t]$ and for {\em all} admissible
input signals in $\sig(V)$ yields the definition of the reachable set:
$$R_I(X_0)=\bigcup_{x\in X_0} \bigcup_{t\in I}\bigcup_{\inp \in
\sig(V)}R(x,\inp,t).$$
Figure~\ref{fig:cont-tree} illustrates the induced trajectories and the reachable states for the case where $X_0=\{x_0\}$.
We will use the same $R_I$ notation also
when $I$ is not an interval but an arbitrary time set.  For example
$R_{[1..r]}(X_0)$ can denote either the states reachable from $X_0$ by a
continuous-time systems at discrete time instants, or states reachable by a
discrete-time system during the first $r$ steps.

\ni

Note that our introductory remark equating the relation between
simulation and reachability computation to the relation between
breadth-first and depth-first exploration of the space of trajectories
corresponds to the commutativity of union:
$$\bigcup_{t\in I}\bigcup_{\inp \in \sig(V)}R(x,\inp,t)=\bigcup_{\inp \in
\sig(V)}\bigcup_{t\in I}R(x,\inp,t).$$

\section{Principles}

In what follows we lay down the principles of one of the most popular
approaches for computing reachable sets which is essentially a set-based
extension of numerical integration.

\subsection{The Abstract Algorithm}

The {\em semigroup property} of dynamical systems, discrete and continuous alike, allows one to compute trajectories incrementally. The reachability operator also admits this property which is expressed as:
$$R_{[0,t_1+t_2]}(X_0)=R_{[0,t_2]}(R_{[0,t_1]}(X_0)).$$
Hence, the computation of $R_I(X_0)$ for an interval $I=[0,L]$ can be
carried out by picking a time step $r$ and executing the
following algorithm:

\newpage
\begin{Algorithm}[Abstract Incremental Reachability]
\label{alg:abst} ~\\
\begin{tabular}{l}
~\\
{\bf Input}: A set $X_0\subset X$\\
{\bf Output}: $Q=R_{[0,L]}(X_0)$ \\
~\\
$P:=Q:=X_0$ \\
{\bf repeat} $i=1,2\ldots$\\
~~$P:=R_{[0,r]}(P)$\\
~~$Q:=Q\cup P$\\
{\bf until} $i=L/r$\\
~\\
\end{tabular}
\end{Algorithm}

\noindent{\bf Remark}: When interested in reachability for {\em unbounded}
horizon, the termination condition $i=L/r$ should be replaced by
$P\subseteq Q$, that is, the newly-computed reachable states are included
in the set of states already computed. With this condition the algorithm is
not guaranteed to terminate. Throughout most of this article we focus on
reachability problems for a bounded time horizon.

\subsection{Representation of Sets}

The most urgent thing needed in order to convert the above
scheme into a working algorithm is to choose a class of subsets of $X$
that can be {\em represented} in the computer and be subject to the
{\em operations} appearing in the algorithm. This is a very important issue,
studied extensively (but often in low dimension) in computer graphics and computational geometry, but
less so in the context of dynamical systems and control, hence we
elaborate on it a bit  bringing in, at least informally, some notions
related to {\em effective computation}.

Mathematically speaking, subsets of $\R^n$ are defined as those points
that satisfy some predicate. Such predicates are {\em syntactic}
descriptions of the set and the points that satisfy them are the {\em
semantic} objects we are interested in. The syntax of mathematics
allows one to define weird types of sets which are not subject to any
useful computation, for example, the set of irrational numbers. In order to compute we need to restrict ourselves
to (syntactically characterized)  classes of sets that satisfy the
following properties:
\begin{enumerate}
\item Every set $P$ in the class $\class$ admits a finite representation.
\item Given a representation of a set $P\in \class$ and a point $x$,
it is possible to check in a finite number of steps whether $x\in P$.
\item For every operation $\op $ on sets that we would like to perform
and every $P_1,P_2\in\class$ we have $P_1 \op P_2\in\class$. Moreover,
given representations of $P_1$ and $P_2$ it should be possible to
compute a representation of $P_1 \op P_2$.
\end{enumerate}
The latter requirement is often referred to as $\class$  being effectively {\em
closed} \comment{\footnote{Not to be confused with topological closure.}} under $\op $. This requirement will later be relaxed into requiring that $\class$ contains a reasonable {\em approximation} of $P_1 \op P_2$. To illustrate these notions, let us consider first a negative example of a class of sets admitting a finite representation but not satisfying requirements 2 and 3 above.  The reachable set of a linear system $\dot{x}=Ax$ can be ``computed'' and represented by a finite formula of the form
$$R_{I}(X_0)=\{x:\exists x_0\in X_0~ \exists t\in I~ x=x_0 e^{At}\},$$
however this representation is not very useful because, in the general case, checking the membership of a point $x$ in this set amounts to
solving the reachability problem itself!  The same holds for checking whether this set intersects another set.  On the other hand, a set
defined by a quantifier-free formula of the form $$\{x:g(x)\geq 0\},$$ where $g$ is some computable function, admits in
principle \comment{Modulo round-off errors and other pathological issues.} a membership check for every $x$: just
evaluate $g(x)$ and compare with $0$.
\begin{figure}
\begin{center}
\scalebox{0.75}{ 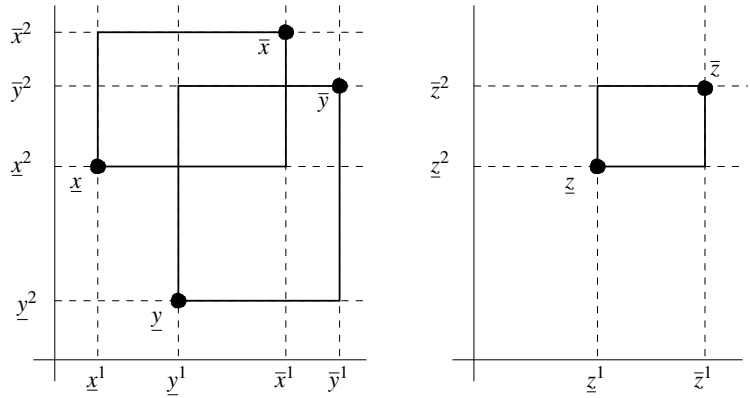 }
\caption{\label{fig:rect} Intersecting two rectangles represented as
$\langle \ul{x},\ov{x}\rangle$ and $\langle \ul{y},\ov{y}\rangle$ to obtain a rectangle
represented as $\langle \ul{z},\ov{z}\rangle$. The computation is done by letting
$\ul{z}^1=\max(\ul{x}^1,\ul{y}^1)$, $\ul{z}^2=\max(\ul{x}^2,\ul{y}^2)$, $\ov{z}^1=\min(\ov{x}^1,\ov{y}^1)$ and $\ov{z}^2=\min(\ov{x}^2,\ov{y}^2)$.}
\end{center}
\end{figure}

As a further illustration consider one of the simplest
classes of sets, hyper-rectangles with axes-parallel edges and rational endpoints. Such a
hyper-rectangle can be represented by its leftmost and rightmost
corners $\ul{x}=(\ul{x}^1,\ldots,\ul{x}^n)$ and
$\ov{x}=(\ov{x}^1,\ldots,\ov{x}^n)$. The set is defined as all points
$x=({x}^1,\ldots,{x}^n)$ satisfying
$$\bigwedge_{i=1}^n \ul{x}^i\leq x^i \leq \ov{x}^i,$$ a condition
which is easy to check. As for operations, this class is effectively
closed under translation (just add the displacement vector to the
endpoints), dilation (multiply the endpoints by a constant) but not
under rotation. As for Boolean set-theoretic operations, it is not
hard to see that rectangles are effectively closed under intersection
by component-wise $\max$ of their leftmost corners and component-wise
$\min$ of their rightmost corners, see Figure~\ref{fig:rect}. However
they are not closed under union and complementation. This is, in fact,
a general phenomenon that we encounter in reachability
computations, where the basic sets that we work with are convex, but
their union is not and hence the reachable sets computed by concrete
realizations of Algorithm~\ref{alg:abst} will be stored as unions
(lists) of convex sets (the recent algorithm of \cite{goran-rajat} is an exception).

As mentioned earlier, sets can be defined using combinations of
inequalities and, not surprisingly, linear inequalities play a
prominent role in the representation of some of the most popular
classes of sets.  We will mostly use {\em convex
polytopes}, bounded polyhedra definable as conjunctions of linear
inequalities. Let us mention, though, that Boolean combinations of
{\em polynomial} inequalities define the {\em semi-algebraic} sets,
which admit some interesting mathematical and computational
results. Their algorithmics is, however, much more complex than that
of polyhedral sets. The only class of sets definable by nonlinear
inequalities for which relatively-efficient algorithms have been developed is the
class of {\em ellipsoids}, convex sets defined as deformations of a unit
circle by a (symmetric and positive definite) linear transformation \cite{ellipsoid-book}.
Ellipsoids can be finitely represented by their center and the
transformation matrix and like polytopes, they are closed under linear
transformations, a fact that facilitates their use in reachability
computation for linear systems. Ellipsoids differ from polytopes by
not being closed under intersection but such intersections can be
approximated to some extent.

\subsection{Convex Polytopes}

In the following we list some facts concerning convex polytopes. These objects, which underlie other domains such as linear programming, admit a very rich theory of which we only scratch the surface. Readers interested in more details and precision \comment{\footnote{For example, we do not bother to make distinctions between strict and non-strict inequalities.}} may consult textbooks such as \cite{schrijver,zeigler}.

A {\em linear inequality} is an inequality of the form $a\cdot x\leq
b$ with $a$ being an $n$-dimensional vector. The set of all points
satisfying a linear inequality is called a {\em halfspace}. Note that
the relationship between halfspaces and linear inequalities is not
one-to-one because any inequality of the form $c a\cdot x\leq c b$,
with $c$ positive, will represent the same set. However using some
conventions \comment{\footnote{For example, insisting that the first non-zero
component of $a$ has absolute value $1$.}} one can establish a unique
representation for each halfspace. A convex polyhedron is an intersection
of finitely many halfspaces. A convex polytope is a
bounded convex polyhedron. A \emph{convex combination} of a set $\{x_1,\ldots,x_l\}$ of points is any
$x=\l_1 x_1+\cdots+\l_l x_l$ such that $$\bigwedge_{i=1}^l \l_i\geq 0 \wedge \sum_{i=1}^l \l_i=1.$$ The
{\em convex hull} of a set $\tp$ of points, denoted by $P=conv(\tp)$, is
the set of all convex combinations of its elements.
Convex polytopes admit two types of canonical representations:
\begin{enumerate}
\item Vertices: each convex polytope $P$ admits a finite minimal set $\tp$ such that $P=conv(\tp)$. The elements of $\tp$ are called the {\em vertices} of $P$.
\item Inequalities: a convex polytope $P$ admits a
minimal set $H=\{H^1,\ldots,H^k\}$ of halfspaces  such that $P=\bigcap_{i=1}^k H^i$. This set is represented syntactically
as a  conjunction of inequalities
$$\bigwedge_{i=1}^k a^i \cdot x \leq b^i.$$
\end{enumerate}
Some operations are easier to perform on one representation and some
on the other. Testing membership $x\in P$ is easier using inequalities
(just evaluation) while using vertices representation, one needs to
solve a system of linear equations to find the $\l$'s. To check whether
 $P_1\cap P_2\not = \emptyset$ one can first combine
syntactically the inequalities of $P_1$ and $P_2$ but in order to
check emptiness, these inequalities should
be brought into a canonical form. On the other hand, $conv(\tp)$ is
always non-empty for any non-empty $\tp$. Various (worst-case exponential) algorithms convert polytopes from one representation to the other.

\comment{
Convex polytopes are bounded polyhedra admitting two dual types of
representations. One as the convex hull of a finite number of vertices
and one as an intersection of a finite number of halfspaces defined by
linear inequalities. [to expand] Some important subclasses of
polytopes will be described later.
Ellipsoids are circles deformed by linear operators [...] and can be
represented by the center and the matrix ...}

The most interesting property of convex polytopes, which is also
shared by ellipsoids, is the fact that they are closed under linear operators,
that is, for a matrix $A$, if $P$ is a convex polytope (resp.\ ellipsoid) so is the set
$$AP=\{Ax:x\in P\}$$ and this property is evidently useful for
set-based simulation.  The operation can be carried out using both
representations of polytopes:  if $P=conv(\{x_1,\ldots,x_l\})$ then
$AP=conv(\{Ax_1,\ldots,Ax_l\})$ and we leave the computation based on
inequalities as an exercise to the reader.

\begin{figure}
\begin{center}
\scalebox{0.85}{ 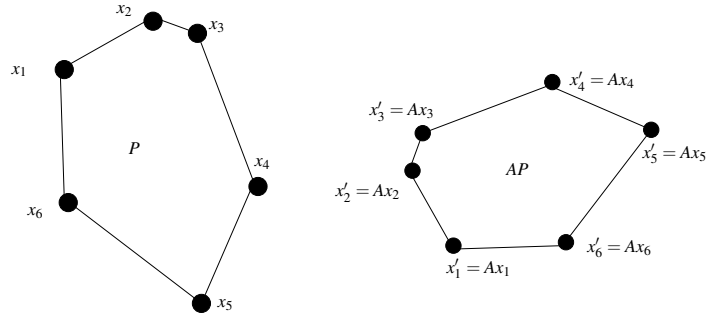 }
\caption{\label{fig:lin-op} Computing $AP$ from $P$ by applying $A$ to the vertices.}
\end{center}
\end{figure}

\section{Linear Systems}

Naturally, the most successful results on reachability computation have been obtained for systems with linear and affine dynamics, that is, systems defined by \emph{linear differential equations}, not to be confused with the simpler ``linear" hybrid automata (LHA) where the derivative of $x$ in any state is independent of $x$. We will start by explaining the treatment of autonomous systems in discrete time, then move to continuous time and then to systems with bounded inputs. In particular we describe some relatively-recent complexity improvements based on a ``lazy" representation of the reachable sets. This algorithm underlies the major verification procedure in the {\bf SpaceEx} tool \cite{sx}  and  can handle quite large systems.

\subsection{Discrete-Time Autonomous Systems}

Consider a system defined by the recurrence equation
$$x_{i+1}=Ax_i.$$
In this case $$R_{[0..L]}(X_0)=\bigcup_{i=0}^L A^i X_0$$
and the abstract algorithm can be realized as follows:

\begin{Algorithm}[Discrete-Time Linear Reachability]~\\
\label{alg:lin:disc}
\begin{tabular}{l}
~\\
{\bf Input}: A set $X_0\subset X$ represented as $conv(\tp_0)$ \\
{\bf Output}: $Q=R_{[0..L]}(X_0)$ represented as a list
 $\{conv(\tp_0),\ldots, conv(\tp_L)\}$  \\
~\\
$P:=Q:=\tp_0$ \\
{\bf repeat} $i=1,2\ldots$\\
~~$P:=AP$\\
~~$Q:=Q\cup P$\\
{\bf until} $i=L$\\
~\\
\end{tabular}
\end{Algorithm}
The complexity of the algorithm, assuming $|\tp_0|=m_0$ is
$O(m_0 L M(n))$ where $M(n)$ is the complexity of matrix-vector
multiplication in $n$ dimensions which is $O(n^3)$ for simple
algorithms and slightly less for fancier ones. As noted, this
algorithm can be applied to other representations of polytopes, to
ellipsoids and any other class of sets closed under linear
transformations. If the purpose of reachability is to detect
intersection with a set $B$ of bad states we can weaken the loop
termination condition into $(i=L) \vee (P\cap B\not = \emptyset)$ where
the intersection test is done by transforming $P$ into an inequalities
representation. If we consider unbounded horizon and want to detect
termination we need to check whether the newly-computed $P$ is included in
$Q$ which can be done by ``sifting'' $P$ through all the polytopes in
$Q$ and checking whether it goes out empty. This is not a simple
operation but can be done. In any case, there is no guarantee that
this condition will ever become true.

\comment{We do not discuss here numerical errors due to the use of floating-point numbers nor the alternative approach based on unbounded-precision rational numbers.}

\subsection{Continuous-Time Autonomous Systems}

The approach just described can be adapted to \emph{continuous-time} systems
of the form
$$\dot{x}=Ax$$ as follows. First it is well known that by choosing a time step $r$ and computing
the corresponding matrix exponential $A'=e^{Ar}$ we obtain a discrete-time system
$$x_{i+1}=A'x_i$$
which approximates the original system in the sense that for every $i$, $x_i$ of the discrete-time system is close to $x[ir]$ of the continuous-time system. The quality of the approximation can be indefinitely improved by taking smaller $r$. We then use the discrete time reachability operator to compute $P'=R_{\{r\}}(P)=A'P$, that is, the successors of $P$ at time $r$ and can use one out of several techniques to compute an approximation of $R_{[0,r]}(P)$ from $P$ and $P'$:

\subsubsection{Make $r$ Small}

This approach, used implicitly by \cite{kurzh}, just makes $r$ small
enough so that subsequent sets overlap each other and the difference
between their unions and the continuous-time reachable set vanishes. \comment{This is the set-based version of the approach which considers the outcome of numerical integration to be the ``real'' continuous-time trajectory.}

\subsubsection{Bloating}

Let $P$ and $P'$ be represented by the sets of vertices $\tp$ and $\tp'$
respectively. The set $\ov{P}=conv(\tp\cup \tp')$ is a good approximation
of $R_{[0,r]}(P)$ but since in general, we would like to obtain an
\emph{over-approximation} (so that if the computed reachable sets does not
intersect with the bad set, we are sure that the actual set does not either)
we can bloat this set to ensure that it is an \emph{outer}
approximation of $R_{[0,r]}(P)$.

This can be done, for example, by pushing the facets of $\ov{P}$ outward by a constant derived from the Taylor approximation of the
curve \cite{linear}. To this end we need first to transform $\ov{P}$ into an inequalities representation. An alternative approach \cite{chutinankrogh99} is to find this over-approximation via an optimization problem. Note that for autonomous systems we can modify Algorithm~\ref{alg:abst} by replacing the initialization by $P:=Q:=R_{[0,r]}(X_0)$ and the iteration by $P:=R_{[r,r](P)}$, that is, the successors after exactly $r$ time units. This way the over-approximation is done only \emph{once} for $conv(\tp_0\cup \tp_1)$ and then $A$ is applied successively to this set \cite{colas-phd}.

\subsubsection{Adding an Error Term}

The last approach that we mention is particularly interesting because
it can be used, as we shall see later, also for non-autonomous
systems as well as nonlinear ones. Let $Y$ and $Y'$ be two subsets of $\R^n$. Their {\em
Minkowski sum} is defined as
$$Y\oplus Y'=\{y+y':y\in Y \wedge y'\in Y'\}.$$ The maximal distance
between the sets $R_{r}(P)$ and $R_{[0,r]}(P)$ can be estimated globally.
Then, one can fix an ``error ball'' $E$ (could be a polytope for
that matter) of that radius and over-approximate $R_{[0,r]}(P)$ as
$AP\oplus E$. Since this computation is equivalent to computing
the reachable set of the discrete time system $x_{i+1}=A'x_i+e$ with
$e\in E$, we can use the techniques for systems with input
described in the next section.

\subsection{Discrete-Time Systems with Input}

We can now move, at last, to open systems of the form
$$x_{i+1}=Ax_i+Bv_i$$
where $v$ ranges over a bounded convex set $V$. The one-step successor of a set
$P$ is defined as
$$P'=\{Ax+Bv: x\in P, v\in V\}=AP\oplus BV.$$
Unlike linear operations that
preserve the number of vertices of a convex polytope, the
Minkowski sum increases their number and its successive application may
prohibitively increase the representation size.
Consequently, methods for reachability under disturbances
need some compromise between exact computation that leads to
explosion, and approximations which keep the representation
size small but may accumulate errors to the point of becoming
useless, a phenomenon also known in numerical analysis as the ``wrapping effect''
\cite{kuhn98,kuhn99}.
We illustrate this tradeoff using three approaches.

\subsubsection{Using Vertices}

\begin{figure}
\begin{center}
\scalebox{0.60}{ 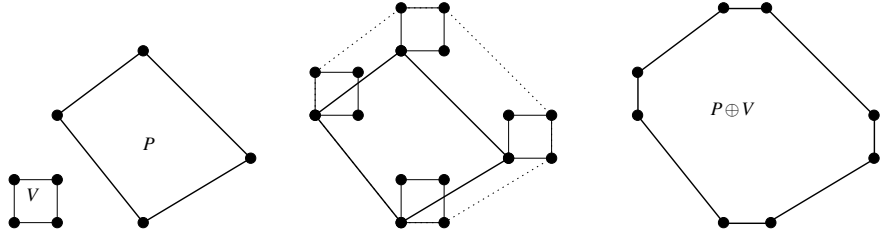 }
\caption{\label{fig:vertexp} Adding a disturbance polytope $V$ to a polytope $P$ leads to a polytope
$P\oplus V$ with more vertices. The phenomenon is more severe in higher dimensions.}
\end{center}
\end{figure}

Assume both $P$ and $V$ are convex polytopes
represented by their vertices, $P=conv(\tp)$ and
$V=conv(\tv)$. Then it is not hard to see that
$$AP\oplus BV =conv(\{Ax+Bv:x\in \tp ,v\in \tv\}).$$ Hence, applying
the affine transformation to all combinations of vertices in $\tp\times \tv$
we obtain all the candidates for vertices of $P'$ (see
Figure~\ref{fig:vertexp}). Of course, not all of these are
actual vertices of $P'$ but there is no known efficient procedure to
detect which are and which are not. Moreover, it may turn out that
the number of actual vertices indeed grows in a super-linear way.
Neglecting the elimination of fictitious vertices and keeping all these points as a
representation of $P'$ will lead to $|\tp|\cdot |\tv|^k$ vertices
after $k$ steps, a completely unacceptable situation.

\subsubsection{Pushing Facets}

\begin{figure}
\begin{center}
\scalebox{0.60}{ 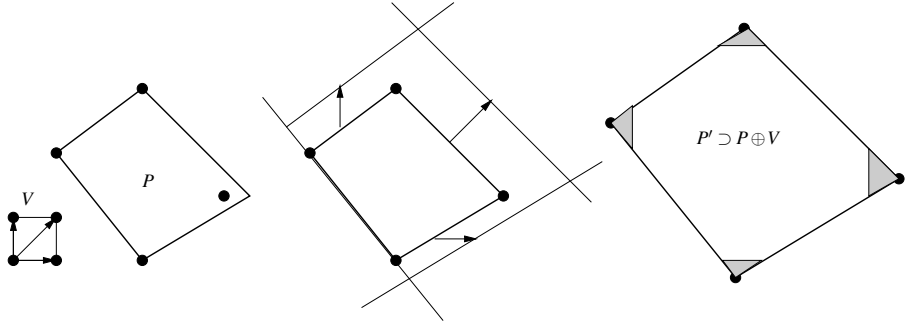 }
\caption{\label{fig:face} Pushing each face of $P$ by the element of $V$ which pushes it
to the maximum. The result will typically  not have more facets or vertices
but may be a proper superset of $P\oplus V$ (shaded triangles represent the over-approximation error).}
\end{center}
\end{figure}

This approach over-approximates the reachable set while keeping its
complexity more or less fixed. Assume $P$ to be represented in (or converted into)
inequality representation. For each supporting halfspace $H^i$ defined
by $a^i \cdot x\leq b^i$, let $v^i \in V$ be the disturbance vector which
pushes $H^i$ in the ``outermost'' way, that is, the one which
maximizes the product $v\cdot a^i$ with the normal to $H^i$.  In the
discrete time setting described here, $v^i\in \tv$ for every $i$.  We
then apply to each $H^i$ the transformation $Ax+Bv^i$ and the
intersection of the hyperplanes thus obtained is an over-approximation
of the successors (see Figure~\ref{fig:face}).

This approach has been developed first in the context of continuous time, starting with \cite{varaiya} who applied it to supporting planes
of ellipsoids and then adapted in \cite{thao-thesis-eng} for polytopes. It is also similar in spirit to the face lifting technique of \cite{face}. In continuous time, the procedure of finding each $v^i$ is a linear program derived from the maximum principle of \emph{optimal control}. Recently is has been demonstrated that by using redundant constraints \cite{thao-redundant} the error can be reduced dramatically for quite large systems.

\subsubsection{Lazy Representation}

The previous sections showed that until recently, the treatment of linear systems with input provided two alternatives: either apply the linear transformation to an ever-increasing number of objects (vertices, inequalities, zonotope generators) and thus slowing down the computation in each step, or accumulate large over-approximation errors. The following observation due to Colas Le~Guernic \cite{colas,zonotope-paper} allows us to benefit from both worlds: it can restrict the application of the linear transformation $A$ to a \emph{fixed} number of points at each step while at the same time \emph{not accumulating} approximation errors. Let us look at two consecutive sets $P_k$ and $P_{k+1}$ in the computation:
$$
{P}_{k}={A}^{k}{P}_0\oplus {A}^{k-1}{BV} \oplus A^{k-2}{BV}\oplus \ldots \oplus
{BV}
$$
and
$$ {P}_{k+1}={A}^{k+1}{P}_0\oplus {A}^k{BV} \oplus A^{k-1}BV \oplus
\ldots \oplus {BV} .$$
As one can see, these two sets ``share'' a lot of common terms that need not be recomputed (and this holds also for continuous-time and variable time-steps). And indeed, the algorithm described in \cite{zonotope-paper} computes the sequence $P_0,\ldots, P_k$ in $O(k(m_0+m)M(n))$ time. From this symbolic/lazy  representation of $P_i$, one can produce an approximating polytope with any desired precision,  but this object is \emph{not} used to compute $P_{i+1}$ and hence the wrapping effect is avoided. \comment{ (see also \cite{sriram-thao-2008}).}

A first prototype implementation of that algorithm could compute the reachable set after $1000$ steps for linear systems with $200$ state variables within $2$ minutes. This algorithm was first discovered for zonotopes (a class of centrally symmetric polytopes closed under Minkowski sum proposed for reachability in \cite{zonotope-antoine}) but was later adapted to arbitrary convex sets represented by \emph{support functions} \cite{support-paper,colas-phd}. A re-engineered version of the algorithm has been implemented into the {\bf SpaceEx} tool \cite{sx}. Although one might get the impression that the reachability problem for linear systems can be considered as solved, the development of {\bf SpaceEx} under the direction of Goran Frehse has confirmed once more  that a lot of work is needed in order to transform bright ideas into a working tool that can robustly handle large non-trivial systems occurring in practice.\ft{Readers are encouraged to download {\bf SpaceEx} at \url{http://spaceex.imag.fr/} to obtain a first hand experience in reachability computation.}

\section{Hybrid and Nonlinear Systems}

Being able to handle quite large linear systems, a major challenge is to extend reachability to richer classes of
systems admitting hybrid or nonlinear dynamics.

\subsection{Hybrid systems}

\begin{figure}
\begin{center}
\scalebox{0.55}{ 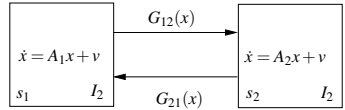 }
\caption{\label{fig:hybrid-aut-example} A simple hybrid system with two modes.}
\end{center}
\end{figure}

The analysis of hybrid systems was the major motivation for developing reachability algorithms because unlike analytical methods, these algorithms can be easily adapted to handle discrete transitions and mode switching. Figure~\ref{fig:hybrid-aut-example} shows a very simple hybrid automaton with two states, each with its own linear dynamics (using another terminology we have here a piecewise-linear or piecewise-affine dynamical system). An (extended) state of a hybrid system is a pair $(s,x)\in S\times X$ where $s$ is the discrete state (mode).
A transitions from state $s_i$ to state $s_j$ may occur when the condition $G_{ij}(x)$ (the transition guard) is satisfied by the current value of $x$. Such conditions are typically comparisons of state variables with thresholds or more generally linear inequalities.
Moreover, while staying at discrete state $s$, the value of $x$ should satisfy additional constraints, known as \emph{state invariants}.\ft{The full hybrid automaton model may also associate transitions with  \emph{reset maps} which are transformations (jumps) applied to $x$ upon a transition.}   Like timed automata, hybrid automata can exhibit dense non-determinism in parts of the state-space where both a transition guard and a state invariant hold. The runs/trajectories of such an automaton are of the form
$$(s_1,x[0])\trans{t_1}  (s_1,x[t_1]) \trans{}(s_2,x[t_1])\trans{t_2}(s_2,x[t_1+t_2])\trans{} \cdots,$$
that is, an interleaving of \emph{continuous trajectories} and \emph{discrete transitions} taken at extended states where guards are satisfied. The adaptation of linear reachability computation so as to compute the reachable subset of $S\times X$ follows the procedure proposed already in \cite{many} for simpler dynamics and implemented in the HyTech tool \cite{hytech}. It goes like this: first, continuous reachability is applied using the dynamics $A_1$ of  $s_1$, while respecting the state invariant $I_1$. Then  the set of reachable states is intersected with the (semantics of the) transition guard $G_{12}$. The outcome serves as an initial set of states in $s_2$ where continuous linear reachability with $A_2$ and $I_2$ is applied and so on.

The real story is, of course, not that simple for the following reasons.
 \begin{enumerate}
 \item The intersection of the reachable states with the state invariant breaks the symbolic lazy representation as soon as some part of the reachable set leaves the invariant. Likewise, the change of dynamics after the transition invalidates the update scheme of the lazy representation and the set has to be over-approximated before doing intersection and reachability in the next state.
\item The dynamics might be ``grazing" the transition guard, intersecting different parts of it at different time steps, thus spawning many subsequent computations. In fact, even a dynamics which proceeds orthogonally toward a single transition guard may spawn many such computations because when small time steps are used, many consecutive sets may have a non-empty intersection with the guard. Consequently, techniques for hybrid reachability such as \cite{colas-antoine-cav-09} tend to cluster these sets together before conducting reachability in the next state thus increasing the over-approximation error. This error can now be controlled using the techniques of  \cite{goran-rajat}.
\item Even in the absence of these phenomena, when there are several transitions outgoing from a state we may end up with an exponentially growing number of runs to explore. 
\end{enumerate}
All these are problems, some tedious and some glorious, that need to be resolved if we want to provide a robustly working tool.

\subsection{Nonlinear Systems}

Many challenging problems in numerous engineering and scientific domains boil down to exploring the behaviors of nonlinear systems. The techniques described so far take advantage of the intimate relationships between linearity and convexity, in particular the identity $A \cdot conv(\tp)=conv(A \tp)$. For nonlinear functions such properties do not hold and new ideas are needed. I sketch briefly two approches for adapting reachability for such systems: one which is general and is based on linearizing the system at various parts of the state-space thus obtaining a piecewise-linear system (a hybrid automaton) to which linear reachability techniques are applied. Other techniques look for more sophisticated data-structures and syntactical objects that can represent sets reached by specific classes of nonlinear systems such as those defined by polynomial dynamics. Unlike linear systems, linear reachability is still in an exploratory phase and it is too early to predict which of the techniques described below will survive.

\subsubsection{Linearization/Hybridization}

\begin{figure}
\begin{center}
\scalebox{0.7}{ 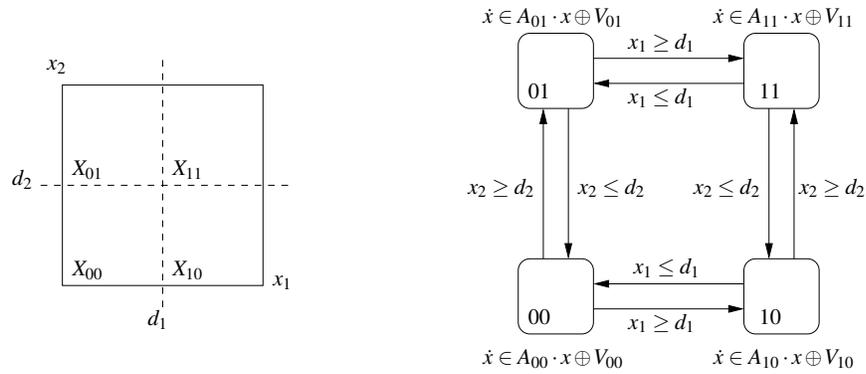 }
\caption{\label{fig:hybridization}Hybridization: a nonlinear system is over-approximated
by a hybrid automaton with an affine dynamics in each state. The transition guards indicate
the conditions for switching between neighboring linearizations.}
\end{center}
\end{figure}

Consider a nonlinear system $\dot{x}=f(x)$ and a partition of its state space, for example into cubes (see Figure~\ref{fig:hybridization}). For each cube $s$ one can compute a linear function $A_s$ and an error polytope $V_s$ such that for every $x\in s$, $f(x)-A_s x \in V_s$ and hence we have a conservative approximation $f(x)\in A_s x\oplus V_s$. We can now build a hybrid automaton (a piecewise-affine dynamical system) whose states correspond to the cubes, and which makes transitions (mode switching) from $s$ to $s'$ whenever $x$ crosses the boundary between them (see Figure~\ref{fig:hybridization}). The automaton provides an over approximation of the nonlinear system in the sense that any run of the latter corresponds to (a projection on $X$ of) a run of the hybrid automaton. This technique, initially developed for doing simulations, has been coined  \emph{hybridization} in \cite{hybridization}. An earlier work \cite{henzinger1998algorithmic} partitions the state-space similarly but approximates $f$ in each cube by a simpler dynamics of the form  $\dot{x} \in C$ for a constant polytope $C$.

\begin{figure}
\begin{center}
\scalebox{1.2}{ 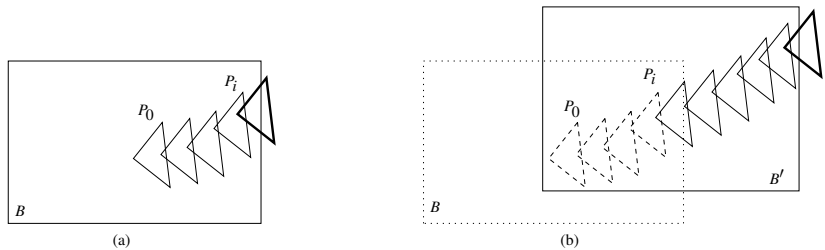 }
\caption{\label{fig:dynamic} Dynamic hybridization: (a) Computing in some
box until intersection with the boundary; (b) Backtracking one
step and computing in a new box.}
\end{center}
\end{figure}

To perform reachability computation on the automaton one can apply the linear techniques described in the preceding section using $A_s$ and $V_s$, as long as the reachable states remain within cube $s$. Whenever some $P_i$ reaches the boundary between $s$ and $s'$ we need to intersect it with the switching surface (the transition guard) and use the obtained result as an initial set for reachability computation in $s'$ using $A_{s'}$ and $V_{s'}$. However, the difficulties previously mentioned concerning reachability for hybrid systems, and in particular the fact that the reachable set may leave a cube and penetrate into exponentially many neighboring cubes, renders hybridization impractical beyond $3$ dimensions. A dynamic version of hybridization, not based on a fixed grid, has been introduced in \cite{reach-nonlin} and is illustrated in Figure~\ref{fig:dynamic}.  The idea is quite simple: first a linearization domain around the initial set is constructed with the corresponding affine dynamics. Linear reachability computation is performed until the first time the computed polytope leaves the domain. Then the last step is undone, and a new linearization domain is constructed around the current set and linear reachability is resumed. Unlike static hybridization,  linearization domains  overlap and do not form a partition, but the inclusion of trajectories still hold by construction (see more details in \cite{reach-nonlin}). The intersection operation and the artificial splitting of sets due to the fixed grid are altogether avoided.

The computation of the linear approximation of $f$ and its error bound are costly procedures and should not be done too often. Hence it is important to choose the size and shape of the linearization domains such that the error is small and the computation stays in each domain as long as possible. This topic has been studied in \cite{thao-curvature} where the curvature of $f$ has been used to construct and orient simplical linearization domains.

\subsubsection{Specialized Methods}

We mention a few other approaches for reachability computation for nonlinear systems.
 For polynomial systems, the Bernstein form of polynomials is used to either enclose the image of a set by the polynomial, or to approximate the polynomial by affine bound functions. The different representations using the Bernstein form include the B\'{e}zier simplices \cite{Dang2006} and the Bernstein expansion \cite{DangTestylierRC2012,DBLP:conf/atva/TestylierD13}.
A nonlinear function can be over-approximated, based on its Taylor expansion,  by a polynomial and an interval which includes all the remainder values. Then, the integration of an ODE system can be done using the Picard operator with reachable sets are represented by boxes \cite{Chen2012}. Similarly, in \cite{Althoff2013} non-linear functions are approximated by linear differential inclusions with dynamical error estimated as the remainder of the Taylor expansion around the current set. As a set representation, ``polynomial zonotopes'' based on a polynomial rather than linear combination of generators are used.

We may conclude that the extension of reachability computation to nonlinear and hybrid systems is a challenging problem which is still
waiting for several conceptual and algorithmic breakthroughs. We believe that the ability to perform reachability computation for
nonlinear systems of non-trivial size can be very useful not only for control systems but also for other application domains such as
\emph{analog circuits} and \emph{systems biology}.  In biological models, uncertainty in parameters and environmental condition is a rule, not an exception, and set-based simulation, combined with other approaches for exploring the uncertainty space of under-determined dynamical systems can be very beneficial.

\section{Related and Future Work}

The idea of set-based numerical integration has several origins. In some sense it can be seen as related to those parts of numerical
analysis, for example \emph{interval analysis} \cite{interval,walter}, that  provide ``robust'' set-based results for numerical computation to compensate for numerical errors. This motivation is slightly different from verification and control where the uncertainty is attributed to an external environment not to the internal computation. Set-based computations also underlie the \emph{abstract interpretation} framework \cite{cousot1977abstract} for static analysis of software.

The idea of applying set-based computation to hybrid systems was among the first contributions of the verification community to hybrid systems research \cite{many} and it has been implemented in the pioneering HyTech tool \cite{hytech}. However this idea was restricted to hybrid automata with very simple continuous dynamics (a constant derivative in each state) where trajectories can be computed without numerical integration. To the best of our knowledge, the first explicit mention of combining numerical integration with approximate representation by polyhedra in the context of verification appeared in \cite{Greenstreet96}. It was recently brought to my attention\ft{G.~Frehse, personal communication.}  that an independent thread of reachability computation, quite similar in motivation and techniques, has developed in the USSR starting from the early 70s \cite{lotov1971}. More citations from this school can be found in \cite{lotov2004interactive}. 
 
The polytope-based techniques described here were developed 
independently in \cite{chutinankrogh99,chutinan-thesis,ChutinanKrogh03} and
in \cite{linear,thao-thesis-eng}. Among other similar techniques that we have not
described in detail, let us mention again the extensive work on
ellipsoids \cite{ellipsoid-book,kurzh,BotchkarevTripakis00,kurzhanski-junior}
and another family of methods \cite{levelset,levelset-more} which uses techniques
such as \emph{level sets}, inspired by numerical solution of partial
differential equations, to compute and represent the
reachable states.
Among the symbolic (non numerical) approaches to the problem let us
mention \cite{sergio} which computes an effective representation of the
reachable states for a restricted class of linear systems.  Attempts
to scale-up reachability techniques to higher dimensions using
compositional methods that analyze an abstract approximate systems
obtained by projections on subsets of the state variables are
described in \cite{GM99} and \cite{projection}.

The interpretation of $V$ as the controller's output rather than disturbance transforms the reachability problem into some open-loop variant of controller synthesis \cite{cfromcs,adversary}. Hence it is natural that the optimization-based approach developed in \cite{morari} for synthesis, has also been applied to reachability computation \cite{morari-bis}. On the other hand, reachability computation can be used to synthesize controllers in the spirit of dynamic programming, as has been demonstrated in \cite{procieee} where a backward reachability operator has been used as part of an algorithm for synthesizing switching controllers.

Another class of methods, called \emph{simulation-based}, for example \cite{jim,girard2006verification,dang2009coverage,sensi,donze2010breach,abbas2013probabilistic}, attempts to obtain the same effect as reachability computation by finitely many simulations, not necessarily of extremal points as in the methods described in this paper. Such techniques may turn out to be superior for nonlinear systems whose dynamics does not preserve convexity and systems whose dynamics is expressed by programs that do not always admit a clean mathematical model.

Alternative approaches to verify continuous and hybrid systems algorithmically attempt to approximate the system by a simpler one, typically a finite-state automaton\ft{In fact, hybridization is another instance of this approach.} \cite{KurshanM91,disc-abst,tiwari2008abstractions}. This can be done by simply partitioning the state space into cubes and defining transitions between adjacent cubes which are connected by trajectories, or by more modern methods, inspired by program verification, such as \emph{predicate abstraction} and \emph{counter-example based refinement} \cite{pred-abst-thao,pred-abst-clarke,ratschan2005safety}. It should be noted, however, that finite-state models based on space partitions suffer from the problem of \emph{false transitivity}: the abstract system may have a run of the form $s_1\to s_2 \to s_3$ while the concrete one has no trajectory $x_1\to x_2\to x_3$ passing through these regions. As a result, the finite-state model will often have too many spurious behaviors to be useful for verification.

Finally let me mention some other issues not discussed so far:
\begin{itemize}
\item Disturbance models: implicit in reachability computation is the assumption that the only restriction on the input signal is that it always remains in $V$. This means that it may oscillate in any frequency between the extremal values of $V$. Such a non realistic assumption may increase the reachable set and render the analysis too pessimistic. This effect can be reduced by composing the system with a bounded-variability non-deterministic model of the input generator but this will increase the size of the system. In fact, the technique of abstraction by projection \cite{projection} does the opposite: it projects away state-variables and converts them into bounded input variables. A more systematic study of precision/complexity tradeoffs could be useful here.
\item Temporal properties: in our presentation we assumed implicitly that systems specifications are simply \emph{invariance} properties, a subclass of safety properties which are violated once a trajectory reaches a forbidden state. It is, of course, possible to follow the usual procedure of taking a more complex temporal property, constructing its automaton and composing with the system model. This procedure will extend the discrete state-space of the system and will make the analysis harder by virtue of having more discrete transitions with the
    usual additional complications associated with detection of cycles in the reachability graph and extraction of concrete trajectories that realize them. 
\item Adaptive algorithms: although we attempt to be as general as possible, it should be admitted that different systems lead to different behaviors of the reachability algorithm, even for linear systems. If we opt for general techniques that do not require a dedicated super-intelligent user, we need to make the algorithms more adaptive to their own behavior and automatically explore different values of their parameters such as time steps, size and shapes of approximating polytopes, quick and approximate inclusion tests, different search strategies (for hybrid models) and more.
\item Numerical aspects: the discourse in this paper treated real-valued computations as well-functioning black boxes. In practice, certain systems can be more problematic in terms of numerical stability or operate in several time scales and this should be taken into account.
\item Differential-algebraic equations: many dynamical systems that model physical phenomena obeying conservation laws are modeled using differential-algebraic equations with relational constraints over variables and derivatives.  In addition to the existing simulation technology for such systems, a specialized reachability theory should be developed, see for example \cite{althoff2013reachability}.
\item Discrete state explosion: the methods described here focus on scalability with respect to the dimensionality of the continuous state-space and tacitly assume that the number of discrete modes is not too large. This assumption can be wrong in at least two contexts: when discrete states are used to encode different parts of a high dimensional state-space as in hybridization or in the verification of complex control software   when there is a relatively small number of continuous variables used to model the environment of the software which has many states. The problem of combining the symbolic representations for discrete  and continuous spaces, for example BDDs and polytopes, is an unsolved problem already for the simplest case of timed automata.
\item Finally, in terms of usability, the whole set-based point of view is currently not the natural one for practitioners (this is true also for verification versus testing in the discrete case) and the extraction of individual trajectories that violate the requirements as well as input signals that induce them will make reachability more acceptable as a powerful model debugging method.
\end{itemize}

 \ni {\bf Acknowledgments}: This work benefitted from discussions with George Pappas, Bruce Krogh, Charles Rockland, Eugene Asarin, Thao Dang, Goran Frehse, Anoine Girard, Alexandre Donz\'{e} and Colas Le Guernic.

\bibliographystyle{eptcs}
\bibliography{all-reach}

\end{document}